\documentclass{emulateapj}
\pdfoutput=1

\catcode`\@=11
\newcommand{\gapprox}{\mathrel{\mathpalette\@versim>}}
\newcommand{\lapprox}{\mathrel{\mathpalette\@versim<}}
\newcommand{\propapprox}{\mathrel{\mathpalette\@versim\propto}}
\newcommand{\@versim}[2]
  {\lower3.1truept\vbox{\baselineskip0pt\lineskip0.5truept
\ialign{$\m@th#1\hfil##\hfil$\crcr#2\crcr\sim\crcr}}}
\catcode`\@=12
\shorttitle{EJECTA OF SUPERNOVA REMNANT N132D}
\shortauthors{BORKOWSKI, HENDRICK, \& REYNOLDS}

\begin{document}
\journalinfo{ApJ Letters}
\submitted{}

\title{X-Ray Emitting Ejecta of Supernova Remnant N132D}

\author{Kazimierz J. Borkowski,\altaffilmark{1}
Sean P. Hendrick,\altaffilmark{2}
\& Stephen P. Reynolds\altaffilmark{1}
}

\altaffiltext{1}{Department of Physics, North Carolina State University,
    Raleigh, NC; kborkow@ncsu.edu.}
\altaffiltext{2}{Department of Physics, Millersville University, 
Millersville, PA.}

\begin{abstract}

The brightest supernova remnant in the Magellanic Clouds, N132D, belongs to 
the rare class of oxygen-rich remnants, about a dozen objects 
that show optical emission from pure heavy-element ejecta.
They originate in explosions of massive stars that produce large amounts of 
O, although only a tiny fraction of that O is found to emit at optical 
wavelengths. We report the detection of substantial amounts of O at X-ray 
wavelengths in a recent 100 ks {\it Chandra} ACIS observation of N132D.
A comparison between subarcsecond-resolution {\it Chandra} and {\it Hubble} 
images reveals a good match between clumpy X-ray and optically emitting 
ejecta on 
large (but not small) scales. Ejecta spectra are dominated by strong lines of 
He- and H-like O; they exhibit substantial spatial variations partially caused 
by patchy absorption within the LMC.
Because optical ejecta are 
concentrated in a 5 pc radius elliptical expanding shell, the detected ejecta 
X-ray emission also originates in this shell. 

\end{abstract}


\section{Introduction}
\label{intro}

The heavy-element ejecta of core-collapse (CC) supernovae (SNe) are
dominated by oxygen. By studying the properties of O and other freshly
synthesized elements in supernova remnants (SNRs), we can infer the
total amount of ejected O, relative abundances, and the spatial
distribution of heavy elements. With this information,
we can then constrain the progenitor main-sequence mass and learn
more about how massive stars explode. Oxygen is most readily seen in
X-rays, as evidenced by a growing number of O-rich SNRs detected by
modern X-ray satellites such as {\it Chandra} and {\it XMM-Newton}.
X-ray detection of plasma enriched in O but deficient in Fe allows us
to identify CC SNRs in the first place.  
High-resolution studies of 
SNRs containing both optical and X-ray--emitting O ejecta 
hold particular promise for advancing our understanding of ejecta
in CC SNRs.

N132D, an X-ray--bright SNR in the Large Magellanic Cloud (LMC),
contains optically emitting O-rich ejecta
\citep{sutherland95,morse95,morse96}.  \citet{morse95} interpreted
their observations in terms of an expanding (1650 km s$^{-1}$)
10 pc ellipsoidal ejecta shell.  Assuming undecelerated
expansion, they estimated the remnant's age at 3150 yr, 10 times older
than Cas A. Spectroscopy with {\it ASCA} \citep{hughes98} and {\it
XMM-Newton} \citep{behar01} showed that the X-ray emission is
dominated by the shocked ambient medium.  X-ray emission from ejecta
has not yet been convincingly demonstrated, although it
might be expected based on the high
(30-35 M$_{\odot}$) mass of the SN progenitor inferred from optical
and UV spectroscopy \citep{blair00}.  We report here the discovery of
clumpy X-ray--emitting O ejecta in N132D with new {\it Chandra} X-ray
observations, matching (on large scales) the optical ejecta morphology
as seen by {\it Hubble}.

\section{Imaging}
\label{imaging}

\begin{figure}
\centering
\includegraphics[scale=0.10]{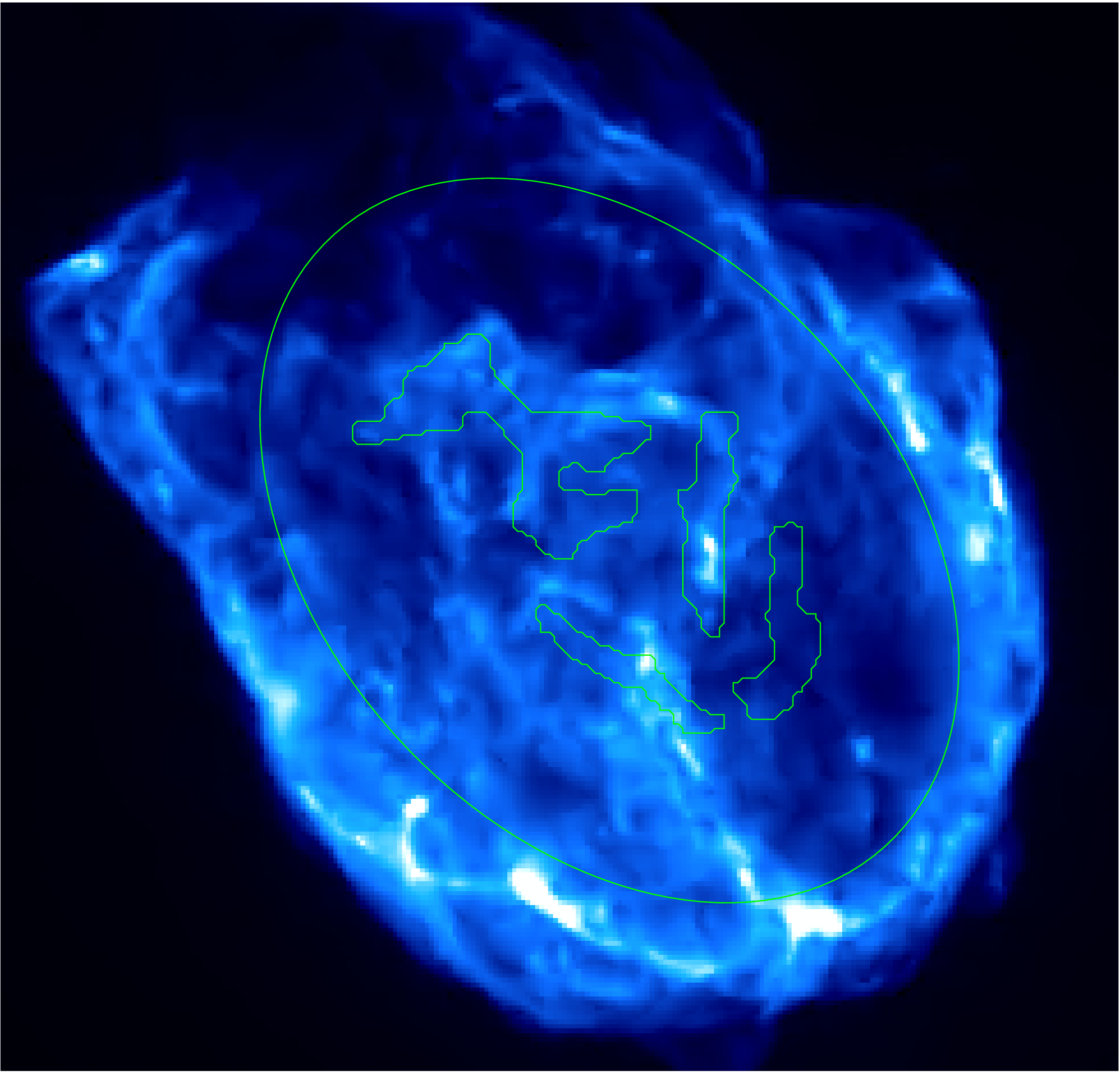}
\includegraphics[scale=0.42]{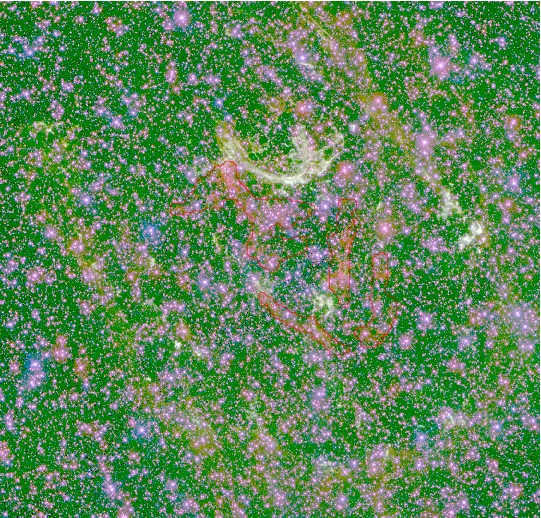}
\caption{Top: {\it Chandra} ACIS image of N132D in the 0.3--7.0 keV energy 
range, smoothed with platelets \citep{willett07}. 
The image is 120'' $\times$ 115'' in size. 
North is up and east is to the left.
Note the complex filamentary 
structure within the remnant's interior (enclosed by an ellipse) and the bright 
outer blast wave. Bottom: A three-color {\it Hubble} ACS image in F475W, F658N, 
and F775W filters (in red, green, and blue, respectively). O-rich 
ejecta are prominent in the F475W filter, while the shocked ISM radiates 
predominantly in the F658N filter (Beasley et al. 2004; see also a high 
quality image in Brown 2007, p.~26). The location of 
optically emitting O-rich ejecta is marked in both images.
\label{images}
}
\end{figure}

N132D was observed by the {\it Chandra X-Ray Observatory} on 2006
January 9, 10, and 16 with the Advanced CCD Imaging Spectrometer (ACIS) 
S3 CCD chip, at the same telescope
roll angle and target location for each observation, for a total
effective exposure of 89.3 ks.  Data were processed with CIAO version 3.4 and
CALDB 3.3.0 with default processing options. While the data were
acquired in the very faint telemetry mode, we chose the default faint
format in data analysis because of a significant loss of source
photons in high surface brightness sections of the remnant with the
very faint format. A large background region covering most of the S3
chip was used for all spectra. Spectral analysis was performed with
XSPEC version 12 \citep{arnaud96}.  We used the nonequilibrium ionization
(NEI) version 2.0 thermal models, based on the APEC/APED spectral codes
\citep{smith01} and augmented by addition of inner-shell processes
\citep{badenes06}.

We obtained $2.8 \times 10^6$ source photons, with about 0.5\%\
background.  Images were smoothed using platelets with a default
smoothing parameter $\gamma = 1/2$ \citep{willett07}; Figure
\ref{images} ({\it top}) shows the remnant's X-ray morphology. We measured the
position of a bright star visible in X-rays on the S3 chip to be
within $0\farcs1$ of its USNO UCAC2 position. 

High spatial resolution imaging of O-rich ejecta was done by the {\it
Hubble Space Telescope} Wide Field Planetary 
Camera 2 \citep[WFPC2;][]{morse96} and the Advanced Camera for Surveys 
\citep[ACS;][]{beasley04}. {\it Hubble} ACS images obtained on 2004 January 22
are most useful, because they provide a complete spatial coverage of
the remnant, and the broad bandpass of the ACS Wide Field Channel (WFC) 
F475W filter
includes [\ion{O}{3}] $\lambda$5007 emission from O ejecta even with
most extreme radial velocities. In addition to the ACS F475W data, we
also use ACS images taken with the F658N and F775W filters, to
highlight the shocked ambient medium emitting strongly in H$\alpha$ and
to distinguish diffuse SNR emission from numerous LMC
stars. Individual ACS exposures (four per filter) were aligned and
combined, and cosmic rays were removed with the on-the-fly
reprocessing system at STScI. We refined the astrometric accuracy of
the final combined images by identifying and measuring positions for a
number of Guide Star Catalog II (GSC2) stars and shifting images from the 
GSC1 reference frame
to the more accurate ($0\farcs3$) GSC2 frame.  We present these images
as a three-color image in Figure \ref{images} ({\it bottom}); red diffuse 
emission marks
the location of O-rich ejecta. 
We outline optically emitting O-rich ejecta, detected both in {\it Hubble} 
and in ground-based data of \citet{morse95}, with solid lines in the 
bottom and 
top panels of Figure \ref{images}.

\begin{figure}
\centering
\includegraphics[scale=0.10]{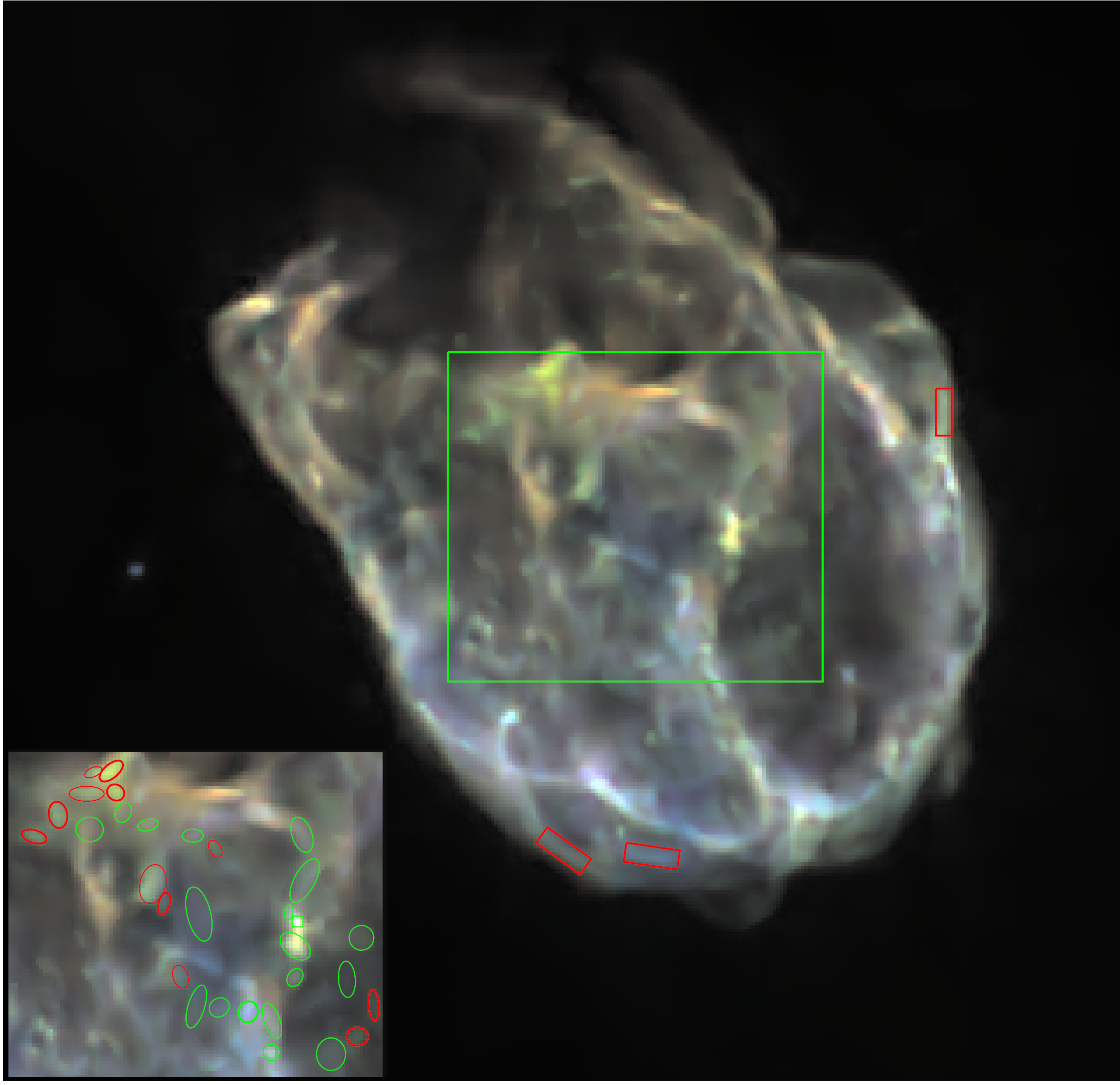}
\caption{Merged image between 0.3 and 7 keV. {\it Red:} 0.3--0.5 keV; 
{\it green:} 0.5--0.75 keV; {\it blue:} 0.75--7 keV. All three images 
were smoothed using 
platelets \citep{willett07}.  X-ray knots and filaments near 
optical O-rich ejecta are shown in the inset. Red ellipses 
show ejecta knots and filaments with particularly strong emission in the 
O band (0.5--0.75 keV).
Regions in green contain 
a mixture of ejecta and the shocked ambient medium. Three selected blast wave
regions are enclosed by red rectangles.  
\label{colorxray}
}
\end{figure}

A similar approach allows us to locate O-rich ejecta in X-rays.  We
constructed a composite three-color image from soft
(0.3--0.5 keV), medium (0.5--0.75 keV), and hard (0.75--7 keV) energy
bands (Fig.~\ref{colorxray}). Emission in the soft band is
particularly sensitive to variations in the interstellar medium (ISM) 
absorption toward
N132D. The medium band contains strong O He$\alpha$ 0.57 keV and O
Ly$\alpha$ 0.65 keV lines; O-rich gas will produce an excess of
emission in this band when compared with the ambient LMC gas. The
broad hard band measures the strength of the overall X-ray emission.
Figure \ref{colorxray} reveals widespread variations in X-ray
colors in the central region. Enhancements in the medium band
({\it greenish knots and filaments}) correlate spatially with optical O-rich
ejecta, implying O-rich X-ray--emitting ejecta with a similar 
large-scale distribution as the optical ejecta.

We attempted to quantify this spectral distinction between knots near
optical O-rich ejecta and emission elsewhere in the interior.  
We selected a large number of X-ray emission enhancements in the N132D
interior (within the ellipse shown in the top panel of Fig.~\ref{images}, 
chosen to exclude the bright outer blastwave). This was
done by repeated thresholding of the image shown in
Figure~\ref{images} ({\it top}), filtered first by the unsharp masking 
procedure,
and further processed with the morphology-based operation ``opening'' (a
basic image-processing algorithm) to spatially separate partly
overlapping emission enhancements.  This procedure
provided us with many spatial regions, preferentially sampling
X-ray knots and filaments, as opposed to more spatially uniform inter-knot
emission.  Our final sample includes 145 regions.

\begin{figure}
\centering
\includegraphics[scale=0.45]{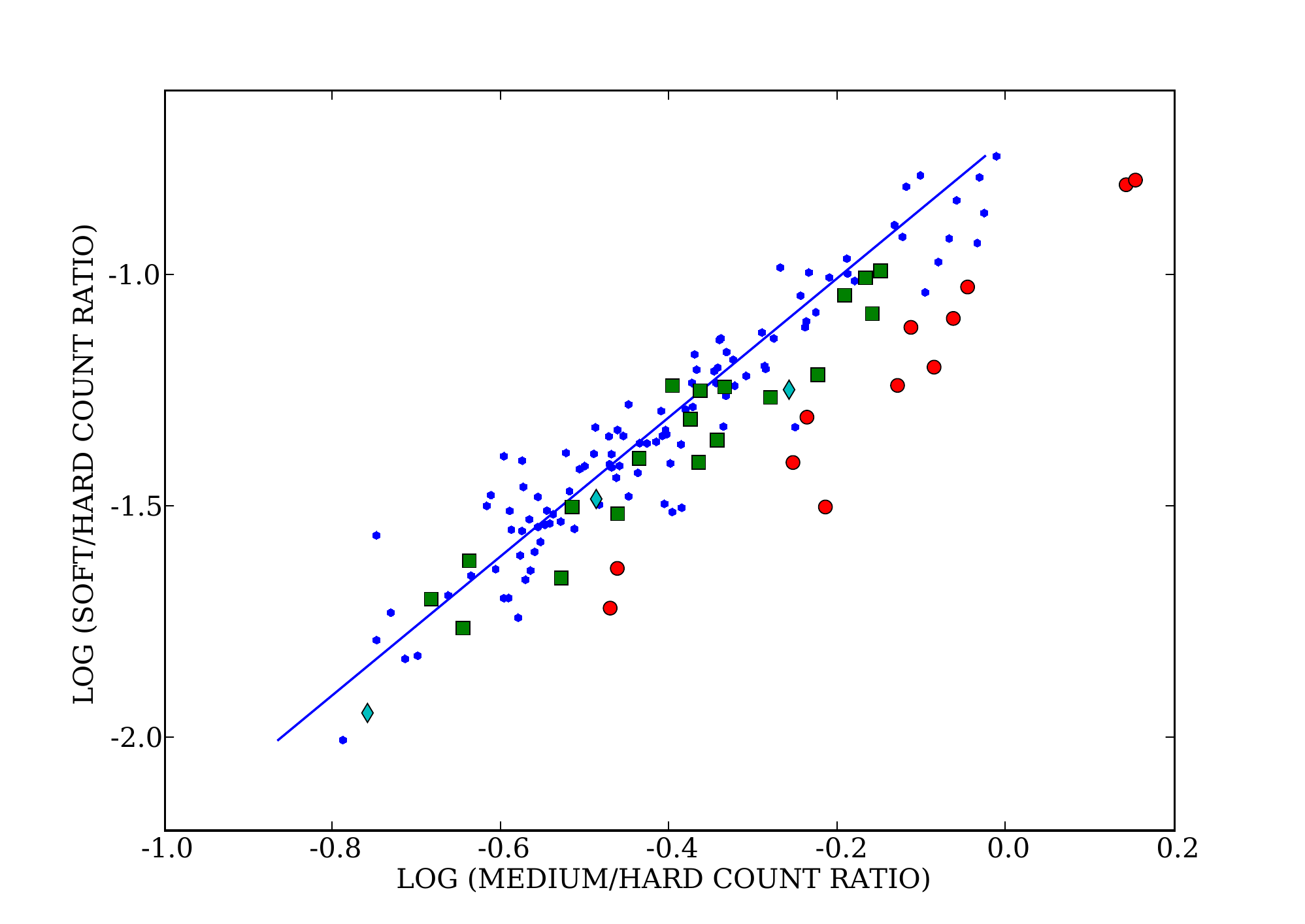}
\caption{Color-color scatter plot, 
0.3--0.5/0.75--7 vs. 0.5--0.75/0.75--7 keV 
count ratios (on log-log scale), for central regions of N132D (within 
the ellipse of Fig.~1, {\it top}). Colors of X-ray--emitting gas not 
overlapping 
spatially with O-rich optical ejecta cluster along a linear regression line 
with slope 3/2, while red and green regions associated with optical ejecta 
(Fig.~2) 
exhibit emission excess in the 0.5--0.75 keV band. Red circles denote 
regions with the most extreme colors, more than $3 \sigma$ away from the 
regression line. Three outer blast wave regions are marked by diamonds.
\label{colorcolor}
}
\end{figure}

We then counted photons within each region in unsmoothed {\it Chandra}
images in each of the three energy bands, arriving at soft ($S$),
medium ($M$), and hard ($H$) counts.  We first examined regions 
not
near O-rich ejecta, by excluding the regions shown in Figure~\ref{images}.
In Figure \ref{colorcolor}, we plot their colors $C_S =
log (S/H)$ and $C_M = log (M/H).$ We also
excluded regions with less than 16 counts (corresponding to a
signal-to-noise ratio $S/N < 4$) in the soft band, in order to reduce
effects of counting (Poisson) noise.  There is a large spread in X-ray
colors in the center of N132D (Fig.~\ref{colorcolor}), but they are
strongly correlated. A linear regression describes this correlation
well, and scatter about the regression line is dominated by the
counting statistics in the soft band. 

We next selected 31 X-ray emission enhancements
located in close proximity to O-rich ejecta; 
they are shown in Figure \ref{colorxray}. Their location in the
color-color plot differs from the other central regions. While their
colors are also strongly correlated, they show excess emission in the
medium band (Fig.~\ref{colorcolor}). Regions with the most extreme
colors appear green in Figure \ref{colorxray}; we marked them by large 
circles in
Figure \ref{colorcolor}. The remaining regions appear blue; Figure
\ref{colorcolor} reveals that they still show an excess emission in
the medium band.  This excess emission indicates X-ray--emitting O-rich
ejecta.  The spatial distributions of optical and X-ray--emitting ejecta
are strongly correlated, although a detailed comparison of Figures
\ref{images} ({\it bottom}) and \ref{colorxray} reveals that there is no exact
correspondence between optical and X-ray emission on the smallest
spatial scales.  Clumpy X-ray--emitting ejecta have the same 
large-scale distribution as optically emitting ejecta, forming an expanding
elliptical shell \citep{morse95}.
 
\section{Spectroscopy}
\label{spectroscopy}

\begin{figure}
\centering
\includegraphics[scale=0.35,angle=90]{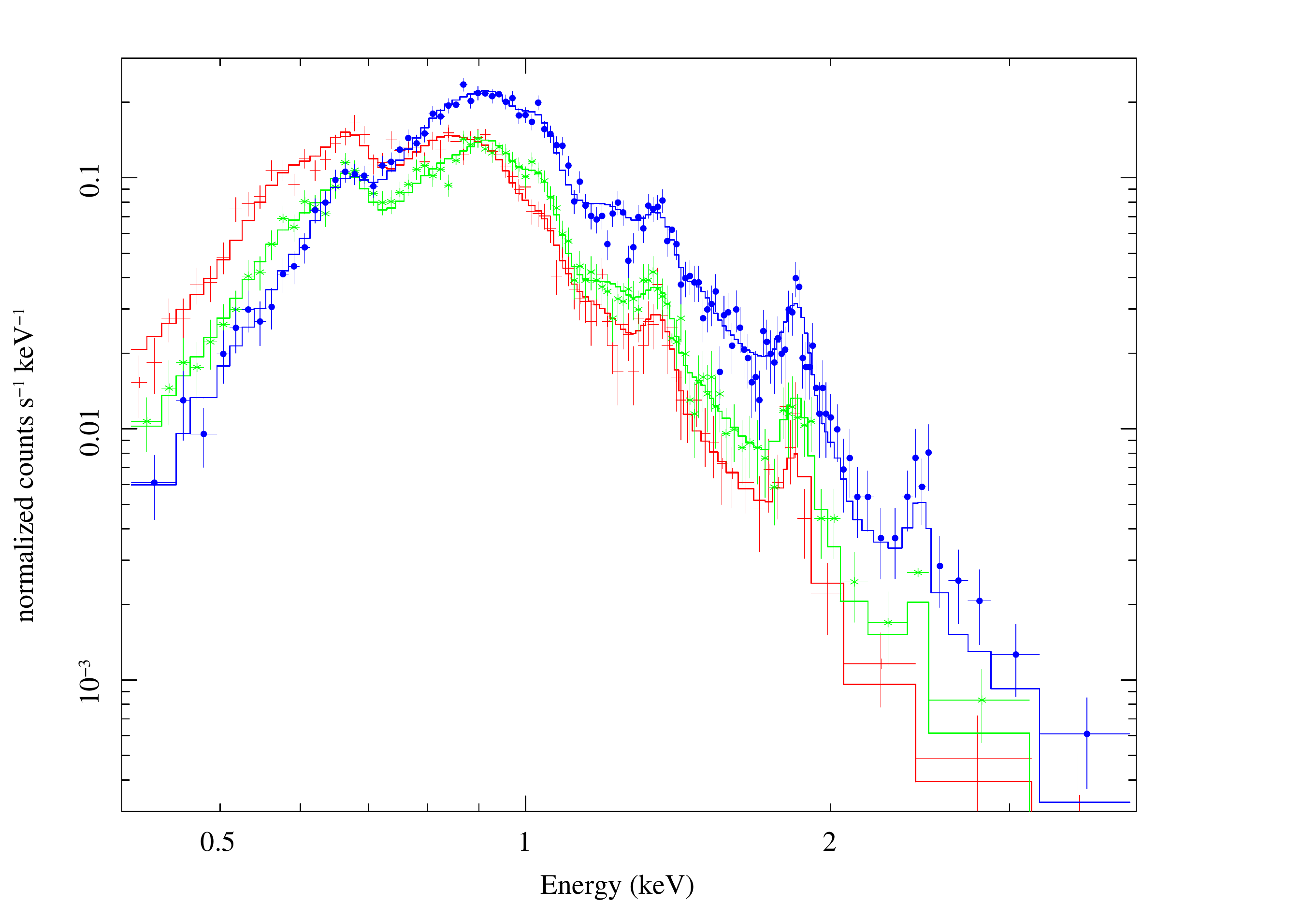}
\caption{Spectra of the blastwave in three locations shown in Fig.~2, one 
in the west ({\it crosses, red}) and two in the south 
({\it stars, green; and filled 
circles, blue}). Plane-shock model 
fits are shown by solid lines. Large spectral variations at low energies are
caused by spatially varying absorption within the LMC. 
\label{ambient}
}
\end{figure}

In order to investigate the large spread in X-ray colors in
N132D, we examined spectra of the blast wave at a number of locations
along its periphery. We found substantial spatial variations in
the ISM absorption, which greatly affect spectra at low photon
energies. Three blast wave spectra in Figure \ref{ambient} demonstrate
this effect.  We modeled them with a plane shock model \citep{borkowski01}, 
using a two-component (Galactic plus LMC) absorption.  The Galactic 
absorption is $5.5
\times 10^{20}$ cm$^{-2}$ at the N132D location \citep[based on H~I
radio observations of][]{stav03}, while the intrinsic absorption $N_H$
within the LMC was allowed to vary. We assumed 0.4 solar abundances
within absorbing LMC material, except for nitrogen, whose abundance is
low (0.1 solar) in the LMC. The same abundances have been assumed for
the emitting gas, except for Ne, Mg, and Fe (with Ni tied to Fe), whose
abundances were allowed to vary. The best-fit value for $N_H$ (LMC) of
$1.4 \times 10^{20}$ cm$^{-2}$ ([0, 4] 90\%\ confidence interval) in
the western blast wave region is consistent with no absorption within
the LMC, while significant absorption, $1.6 (1.2, 2.1) \times 10^{21}$
cm$^{-2}$ and $4.1 (3.6, 4.7) \times 10^{21}$ cm$^{-2}$ is required in
the two southern regions. Plasma temperatures are nearly equal, 0.66,
0.71, and 0.67 keV, respectively---while ionization ages are longer in
the south [$6.1 (3.8, 8.4) \times 10^{11}$ cm$^{-3}$ s and $15 (8.7,
23) \times 10^{11}$ cm$^{-3}$ s] than in the west [$4.0 (2.9, 5.3)
\times 10^{11}$ cm$^{-3}$ s].

The X-ray colors of the blastwave depend strongly on $N_H$; colors
corresponding to spectra of Figure \ref{ambient}, and plotted in
Figure \ref{colorcolor}, spread along the regression line.
Colors become bluer down the line as absorption increases. We conclude
that spatial variations in interstellar absorption are mostly
responsible for the large range in X-ray colors seen in Figure
\ref{colorxray}. The significant difference in $N_H$ between the two
adjacent regions in the south demonstrates that absorption is clumpy
on small spatial scales, although a large-scale north-south absorption
gradient is also present.  \citet{banas97} argued that N132D is
physically associated with a molecular cloud located $\sim 1\farcm 5$
south of the remnant. While there is little spatial overlap between
X-ray, CO, and H$_2$ emission \citep{banas97,tappe06}, Australia Telescope 
Compact Array \ion{H}{1}
21 cm survey images \citep{kim03} reveal the presence of substantial
amounts of \ion{H}{1} emission around this molecular cloud, extending
to the northern tip of N132D. It is likely that this gas just outside
the molecular cloud is responsible for the patchy X-ray absorption.

\begin{figure}
\centering
\includegraphics[scale=0.35,angle=0]{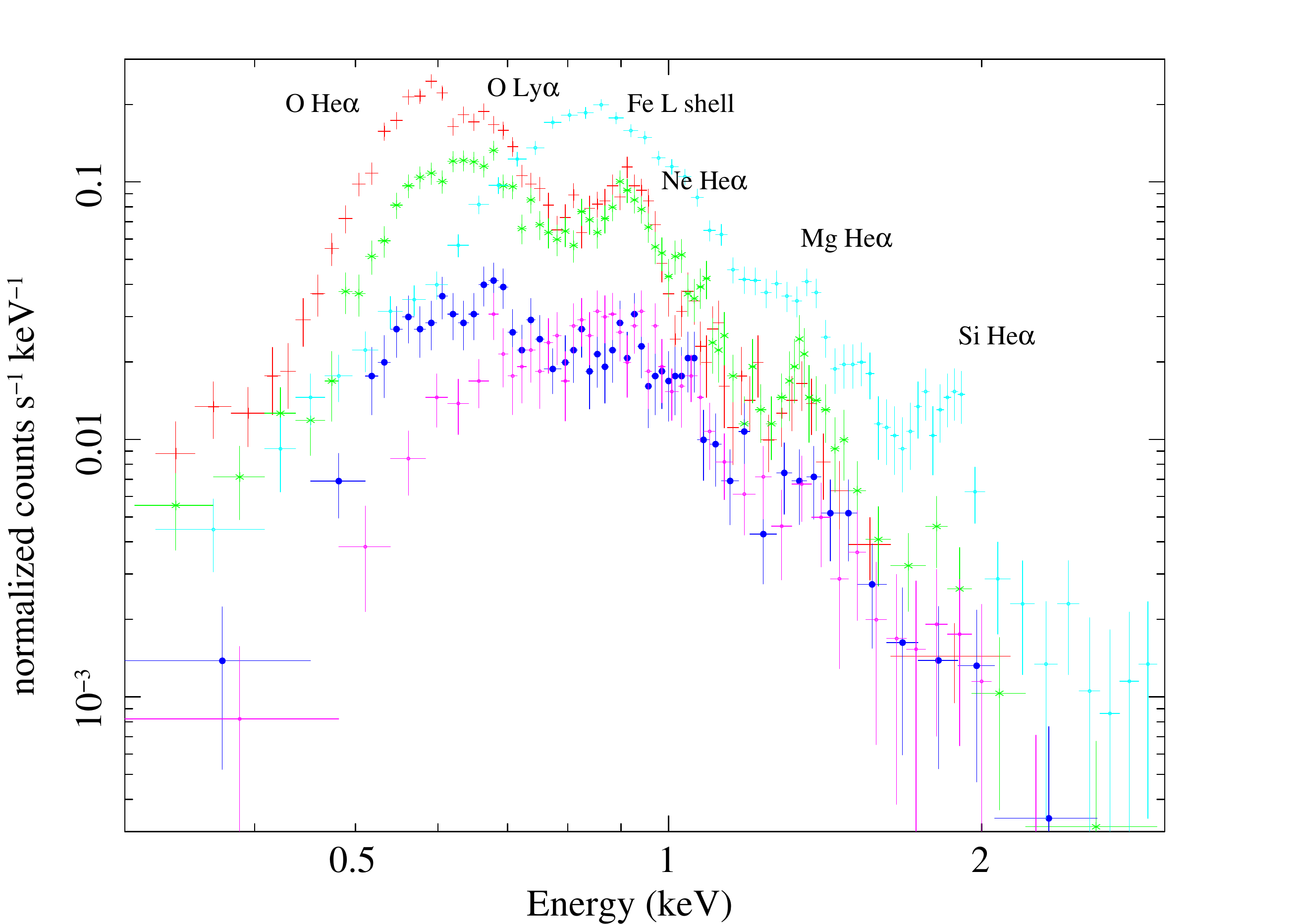}
\caption{X-ray spectra of ejecta in four locations shown in Fig.~2 by thick 
red ellipses (we added spectra from adjacent ellipses in order to improve 
S/N ratio). For comparison, an X-ray spectrum of the shocked ambient gas is 
also shown (Fig. 2, {\it thick green ellipse}). {\it From top to bottom} 
(at 0.6 keV): North, east, ambient, west, and central spectra, with 
colors ($C_M$, $C_S$) equal to (0.15, -0.80), (-0.10, -1.14), (-0.64, -1.62), 
(-0.17, -1.34),
and (-0.47, -1.72), respectively. Prominent X-ray lines are labeled. Oxygen 
lines in ejecta spectra are stronger than in the blast wave spectrum.
\label{oxygenejecta}
}
\end{figure}

Spectra of several regions associated with optically emitting O-rich
ejecta and with excess emission in the medium band are shown in Figure
\ref{oxygenejecta}. These regions are marked by thick red ellipses
in Figure \ref{colorxray}; we added spectra for adjacent regions if
their colors were similar. The top spectrum corresponds to two
northern knots with extremely soft hardness ratios (these knots are
located in the top right corner of Figure \ref{colorcolor}). Bright
optically emitting O-rich ejecta filaments overlap spatially
\citep{morse95} at the location of the northern knots. Strong O lines
dominate the X-ray spectrum; they are much stronger than O lines in
the blast wave spectra of Figure \ref{ambient}. The O He$\alpha$ line
of the He-like O$^{+6}$ ion is stronger than the O Ly$\alpha$ line of
the hydrogenic O$^{+7}$ ion. It is likely that a lower than average
ionization state of O accounts for the softness of the X-ray
spectrum. Spectra of ejecta in the east and west, corresponding to the 
blueshifted
filaments B4 and B1 of \citet{morse95}, also show strong O lines.
These are most typical of ejecta spectra in N132D. The O Ly$\alpha$
line is now stronger than the O He$\alpha$ line, resulting in harder
spectra and bluer X-ray colors.  The X-ray colors of the eastern knots
place them in the main ``clump'' of ejecta knots in the X-ray
color-color plot (Fig.~\ref{colorcolor}), significantly below the
color-color relationship for the shocked ambient gas. 
The X-ray
spectrum of a small central ejecta knot (the bottom spectrum in
Fig.~\ref{oxygenejecta}) corresponds to even bluer X-ray colors ($C_M
= -0.47$, $C_S = -1.72$). A comparison with the hard ($C_M
= -0.64$, $C_S = -1.61$) spectrum of a bright interior knot with the 
normal (LMC) abundance (Fig.~\ref{oxygenejecta}) reveals the presence of 
stronger than average O lines, but they are weaker than in other ejecta 
spectra, likely
because of the increased absorption.  A substantial amount of
absorbing LMC gas appears to be present toward the very center of
N132D; a central ``blue hole'' seen in Figure \ref{colorxray}
corresponds to an absorbing ISM cloud with $N_H ({\rm LMC}) = 3 \times
10^{21}$ cm$^{-2}$.

Strong O lines in ejecta spectra may be due either to enhanced O
abundances or to short ionization ages. We fit the top spectrum of
Figure~\ref{oxygenejecta} with a plane shock model with temperature of
0.59 keV and ionization age of $4 \times 10^{10}$ cm$^{-3}$ s and
with nearly the same abundances as in the blast wave. For electron
densities of $> 15$ cm$^{-3}$ \citep[typical of the N132D blast
wave;][]{williams06}, the very short ionization age implies a shock
age of less than 100 yr. This is clearly an unreasonably short shock
age for a 3000 yr old SNR. Problems arise also if low ($\sim 1$
cm$^{-3}$) electron densities are assumed, as such low-density gas
with LMC abundances is too faint in X-rays.
This would also require a
substantial (an order of magnitude or more) pressure imbalance between
the ejecta and the blast wave, rather difficult to sustain for a
prolonged period of time. These severe difficulties suggest that the
alternative explanation of strong O lines in terms of O-rich ejecta is
the correct one. The determination of plasma conditions and abundances
within the ejecta would require a separate, detailed spectroscopic
analysis involving a superposition of heavy-element ejecta and blast 
wave emission. 
We just note here that the low O Ly$\alpha$/O He$\alpha$
line ratio of 0.41 (0.36 after correction for absorption) in the plane
shock fit just mentioned implies a low (0.2 keV) plasma temperature
under conditions of collisional ionization equilibrium expected
in dense shocked O-rich ejecta clumps. It is likely the observed
clumpy ejecta emission arises in relatively dense O-rich ejecta, which
have been shocked by low-velocity shocks propagating into O-rich
clumps. We estimate that a few tenths of $M_\odot$ of O is present in
these clumps.

\section{DISCUSSION}
\label{summ}

High spatial resolution imaging and spectroscopy of N132D with {\it Chandra} 
reveal the presence of clumpy O-rich ejecta in its center. The spatial 
distribution of X-ray--emitting ejecta correlates well with optically emitting 
ejecta on large (but not small) scales. This shows that optical and X-ray 
emission can be closely linked in dynamically advanced remnants such as N132D, 
where most of the O might have already been shocked. In N132D, ejecta are 
located in an expanding ellipsoidal shell.
No such shell is present in either G292.0+1.8 
or Pup A, two O-rich Galactic SNRs with ages similar to N132D. 
Optically emitting ejecta in G292.0+1.8 are distributed throughout much of the 
remnant, with little detailed correlation with X-rays \citep{winkler06}. 
Asymmetrically distributed optical ejecta filaments in Pup A are moving in 
the opposite direction to its neutron star \citep{winkler07}.
This demonstrates that the innermost metal-rich ejecta are strongly affected 
by poorly understood processes leading to CC explosions. No correlation of 
optical emission with more evenly distributed X-ray--emitting ejecta has been 
reported.
X-ray expansion velocities in N132D are poorly known;
\citet{hwang93} found 
(marginal) evidence for an X-ray expansion consistent with the optical 
expansion in spatially integrated 
spectra obtained by {\it Einstein}, but archival {\it Chandra} and 
{\it XMM-Newton} grating spectra are more suitable for studying the ejecta 
kinematics. 
The origin of the X-ray and optically emitting ejecta shell is unknown at this 
time. It may be the location of a reverse shock propagating into ejecta; 
more likely, it may be an 
ejecta shell created by a Ni bubble effect shortly after the SN explosion, 
similar to what is seen in the O-rich SNR B0049-73.6 \citep{hendrick05}. 
More detailed analysis of existing {\it Chandra} and 
{\it XMM-Newton} observations is warranted.  

\acknowledgments We thank 
Paul Plucinsky for help planning the {\it Chandra}
observations. 
This work was supported 
by NASA through grants SAO G05-6053A and SAO G05-6053B.

\end{document}